\documentclass[10pt,leqno]{amsart}
\usepackage{graphicx}
\usepackage{indentfirst,csquotes}

\usepackage{subcaption}
\usepackage{amssymb,amsthm,amsmath}
\usepackage{xcolor,paralist,hyperref,titlesec,fancyhdr,etoolbox}
\usepackage{secdot}

\definecolor{myyellow}{rgb}{1,1,0.6} % A slightly softer yellow
\definecolor{myblue}{rgb}{0.7,0.9,1}  % A light blue

\titleformat{\section}[display]{\normalfont\huge\bfseries\centering}{\thesection}{10pt}{\Large}
\titlespacing*{\section}{0pt}{0ex}{0ex}

\hypersetup{ colorlinks=true, linkcolor=black, filecolor=black, urlcolor=black }

\usepackage{lipsum}

\begin{document}
\title{Non-destructive diagnostics of fiber orientation in large planar fiber-reinforced concrete specimens} %%%%%%%%%%%%
\author{Václav Papež, Karel Künzel, Petr Konrád, Kristýna Carrera and Petr Konvalinka}
\date{\today}

\email{petr.konrad@fsv.cvut.cz}
\maketitle
\pagestyle{plain}

\let\thefootnote\relax

\begin{abstract}

\end{abstract} %%%%%%%%%

\bigskip

\noindent The mechanical performance of fiber-reinforced concrete is critically dependent on the orientation and concentration of its reinforcing fibers. This paper presents a non-destructive method for diagnosing fiber parameters in large planar specimens. The proposed technique utilizes an electromagnetic measurement system using electrical coils with commercial LCR meters, but also a novel custom-made meter suitable for factory environments, offering a practical solution for quality control in prefabrication of high-performance structural elements. This study details the theoretical background, experimental setup and methodology, and provides an evaluation of the diagnostic system's effectiveness and accuracy. The results demonstrate a strong correlation between the electromagnetic measurements and the actual fiber parameters, confirming the method's reliability.

$\,$

$\,$

\section{Introduction}
\label{intro}

Concrete is one of the most commonly used structural materials worldwide \cite{Miller2020}. It offers good compressive strength, but poor tensile strength, which is negated by the use of steel rod reinforcement \cite{Mosley2007}. Another method of reinforcement is the use of fibers of various materials, but also very commonly steel. fibers are used to increase the toughness of concrete, to mitigate its otherwise brittle nature, to control the formation and spread of cracks in applications ranging from industrial floors to precast elements for bridge construction \cite{Bentur2007}. In certain applications, fibers can even replace the role of reinforcing bars, mainly in elements not requiring high loads \cite{DiPrisco2009}. fibers are added during mixing of fresh concrete, which means their distribution and orientation is semi-random \cite{BOULEKBACHE20101664}. This is a problem if fibers are to be used efficiently in structural elements where majority of fibers should work to reinforce the concrete in certain direction, i.e. against the designed load. Orienting the fibers as needed is, therefore, desirable. One such method is the magnetic orientation. Fresh fiber-reinforced concrete is exposed to a strong magnetic field, where ferromagnetic steel fibers turn parallel to the magnetic field lines. This method has already been extensively studied \cite{WIJFFELS2017342,Ghailan02012020,ALRIFAI2024134796,cryst11070837} also by the research team in previous studies \cite{CARRERA2023104534,acta1,doi:10.1061-ASCE-MT.1943-5533.0004342}. Clearly, as concrete is opaque and the fiber orientation's success cannot be visually checked, the orientation research must go hand in hand with an ability to non-destructively assess the orientation. Also, such assessment is necessary for fiber-reinforced elements made without any orientation process applied as a form of quality control, since the randomness of fiber orientation and concentration is a major obstacle in guaranteeing mechanical performance of the finished structure. 

One way to see each fiber inside the concrete volume would be X-ray scanning. This method however is not practical for quick quality control of large elements, although it can be used in laboratory studies. Most common, and probably the most practical, are methods based on magnetic interactions between the ferromagnetic fibers and a measuring device -- an electrical coil. In literature, there can be found studies focusing on inductance measurements \cite{Ferrara2012}. In the authors' own previous research, measurements of the coil's quality factor was done instead, as it offers higher sensitivity to the presence of steel fibers and their orientation. The basis of such measurement is the use of a high quality coil with a high quality factor over a wide frequency range. A concrete specimen was inserted inside the measuring air coil. Concrete does not alter the coil's properties, but steel of the fibers does. Using this approach, it was possible to measure the level of fiber orientation in the direction of the coil's axis using reference non-oriented and oriented specimens. A specimen with successfully oriented fibers and a specimen with a higher fiber content decreases the quality factor of the measuring coil. An apparent drawback of this method is the use of relatively small specimens and/or very large measuring coils, while the coil only "summarizes" fiber parameters in the entire volume currently inside the coil.

A logical step forward in this principle of non-destructive measurement is to develop a one-surface-only device, that would not require insertion of the measured volume inside it. This would open the possibilities of quality-checking commercially available precast concrete elements with unlimited lengths and widths. On the other hand, the device still must preserve the required sensitivity using the coil's quality factor. Design and testing of such device is the topic of this study. The paper is structures as follows. First, the theoretical background for using the quality factor is presented. Next, still using the air coil with inserted specimens, the theoretical assumptions are confirmed. After that, the new half-toroidal coil for surface measurements is presented with pilot experiments to asses its functionality. In the end, conclusions are drawn with a future outlook.

\section{Theoretical background}
\subsection{The quality factor}

\begin{figure}
\centering
\includegraphics[width=0.3\textwidth]{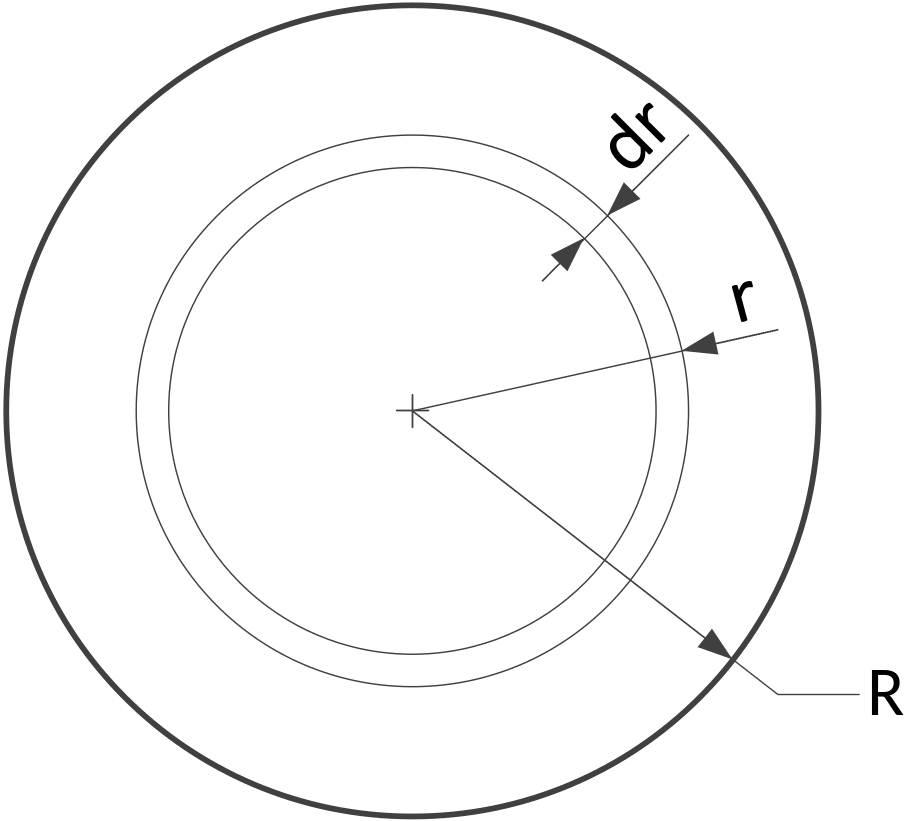}
\caption{Cross-section of a fiber with diameter $R$ with indicated current path with diameter $r$ and thickness $dr$.}
\label{fig:fibercross}
\end{figure}

We will analyze the influence of a fiber on the quality factor of a measuring coil using the perturbation method. The procedure is based on knowing the magnetic field in the coil without inserted fibers and on the assumption that inserting the fibers will only negligibly change the field in the volume outside the fibers. In conductive fibers inserted into the alternating magnetic field of the measuring coil, eddy currents will be induced by the magnetic field, causing electrical power absorption within the fiber volume. This power will manifest as a damping resistance at the coil's terminals, which will increase its damping and decrease its quality factor. In a fiber inserted into an axial magnetic field, the induced current will flow along current paths in the shape of concentric circles, distributed across the fiber's cross-section (Figure \ref{fig:fibercross}).

The effective voltage $U_{RMS}$ induced on a current path with radius $r$ is then given as

\begin{equation}
    U_{RMS} = H_{RMS} \pi \mu \omega r^2
\end{equation}

where $H_{RMS}$ is the effective value of the magnetic field intensity, $\mu$ is the medium's permeability and $\omega$ is the circular working frequency.

The effective current $I_{RMS}$ passing through a current path with the length of $l$ and radius $r$ is given as

\begin{equation}
    I_{RMS} = l\frac{U_{RMS}}{2\pi r \rho} dr = H_{RMS} \frac{\mu \omega l r}{2 \rho} dr = \frac{1}{2} H_{RMS} \sigma \mu \omega l r dr
\end{equation}

where $\rho$ is the electrical resistivity and its reciprocal value $\sigma$ is electrical conductivity of the fiber material. The power $P$, which is absorbed by the current path with the length of $l$ and radius $r$ can be expressed as a product of current $I_{RMS}$ and voltage $U_{RMS}$ and written as

\begin{equation}
    dP = \frac{1}{2} H_{RMS}^2 \sigma \pi \mu^2 \omega^2 l r^3 dr
\end{equation}

If a fiber with constant material characteristics is put into the magnetic field with constant effective value of the magnetic field intensity $H_{RMS}$, then the power $P$ absorbed in the whole volume of the fiber can be expressed as integrating along this entire volume

\begin{equation}
    P = \int_{0}^{R} dP dr
\end{equation}

which can be expanded as follows

\begin{equation}
    P = \frac{1}{2} H_{RMS}^2 \sigma \pi \mu^2 \omega^2 l \int_{0}^{R} r^3 dr = \frac{1}{8} H_{RMS}^2 \sigma \pi \mu^2 \omega^2 l R^4
\end{equation}

If we consider measurements at higher frequencies, where the skin effect influences the magnetic field penetrating the fiber, the magnetic field intensity within the fiber will not be constant. The intensity $H_{RMS}$ within the fiber will only be equal to $H_{RMS}$ in the surrounding environment in a thin surface layer. With increasing distance from the surface, it will decrease exponentially with an attenuation corresponding to the equivalent penetration depth $\delta$

\begin{equation}
    \delta = \sqrt{\frac{2}{\omega \mu \sigma}}
\end{equation}

and the intensity of the magnetic field inside the fiber will then be given as

\begin{equation}
    H(r) = H_0 e^{\frac{r-R}{\delta}}
\end{equation}

where $H_0$ is the intensity on the fiber's surface. The skin effect will cause decreased absorbed power $P_{skin}$ according to the following formula

\begin{equation}
    P_{skin} = H_{RMS}^2 \sigma \pi \mu^2 \omega^2 l \int_{0}^{R} r^3 e^{2\frac{r-R}{\delta}} dr
\end{equation}

Solving the integral using per partes method leads to the expression

\begin{equation}
    P_{skin} = H_{RMS}^2 \sigma \pi \mu^2 \omega^2 l \left( \frac{\delta}{2} R^3 - \frac{3}{4} \delta^2 R^2 + \frac{3}{4} \delta^3 R - \frac{3}{8} \delta^4 + \frac{3}{8} \delta^4 e^{-\frac{2R}{\delta}} \right)
    \label{eq:skin}
\end{equation}

For the penetration depth lower than the fiber's radius, the absorbed power can be approximately expressed as power absorbed in a surface layer with equivalent depth at constant magnetic field intensity as

\begin{equation}
    P_{skin} = H_{RMS}^2 \sigma \pi \mu^2 \omega^2 l \int_{R-\delta}^{R} r^3 dr = \frac{1}{8} H_{RMS}^2 \sigma \pi \mu^2 \omega^2 l (R^4 - (R-\delta)^4)
    \label{eq:small}
\end{equation}

The influence of absorbed power on the quality factor of the measuring coil can be determined by the magnitude of the damping resistance, which is how the absorbed power manifests at the coil's terminals. Additionally, the axial voltage and magnetic field intensity inside the coil are related according to

\begin{equation}
    U = H \mu \omega SN
\end{equation}

where $S$ is the coil's cross-section and $N$ number of turns. We can rewrite the equation for the magnetic field intensity $H$ and substitute it in equation \ref{eq:skin} as follows

\begin{equation}
    H = \frac{4U}{\pi D^2 \mu \omega N}
\end{equation}

\begin{equation}
    P_{skin} = \frac{U_{RMS}^2 \sigma l}{\pi D^4 N^2} \left( 8 \delta R^3 - 12 \delta^2 R^2 + 12 \delta^3 R - 6 \delta^4 + 6 \delta^4 e^{-\frac{2R}{\delta}} \right) = \frac{U_{RMS}^2}{R_d}
\end{equation}

Or similarly we can simplify the equation \ref{eq:small} for small depth of penetration

\begin{equation}
    P_{skin} = 2 U_{RMS}^2 \frac{\sigma l}{\pi D^4 N^2} (R^4 - (R-\delta)^4) = \frac{U_{RMS}^2}{R_d}
\end{equation}

where $R_d$ is the damping resistance at the coil's terminals which represents the losses inside the fiber. $R_d$ could also be rewritten as

\begin{equation}
    R_d = \frac{\pi D^4 N^2}{2 \sigma l \left( 4 \delta R^3 - 6 \delta^2 R^2 + 6 \delta^3 R - 3 \delta^4 (1+e^{-\frac{2R}{\delta}}) \right)}
\end{equation}

or for small depth of penetration as

\begin{equation}
    R_d = \frac{\pi D^4 N^2}{2 \sigma l (R^4 - (R-\delta)^4)}
\end{equation}

Difference between these expressions for $R_d$ is for depths of penetration $\delta$ lower than 0.2$R$ below 3\%. All of the above applies only to a case where the fiber's longitudinal axis is aligned to the coil's axis. If that is not the case, the magnetic field intensity in the fiber's axis will not be equal to the coil's magnetic field intensity. 

\begin{figure}
\centering
\includegraphics[width=0.7\textwidth]{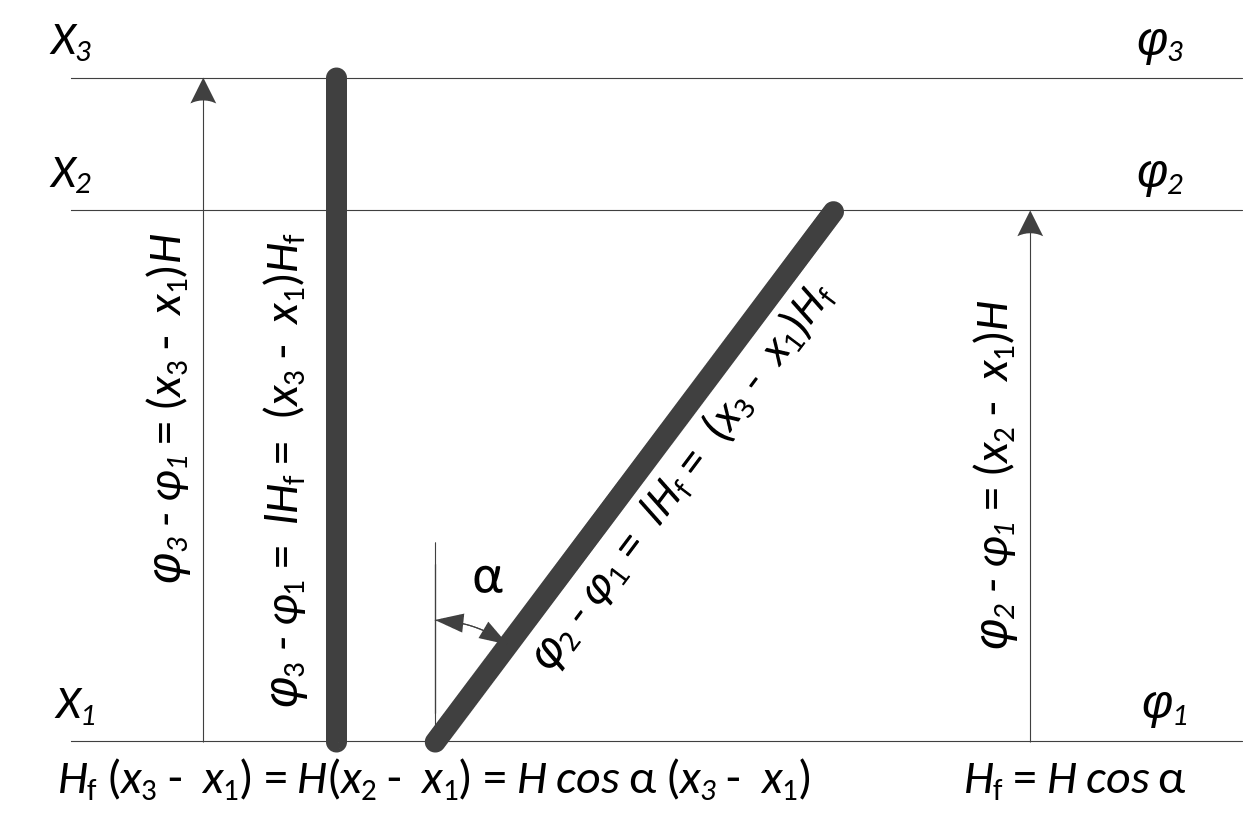}
\caption{Intensity of the magnetic field in the fiber according to linearly increasing magnetic potential, where $x$ is a coordinate parallel to the magnetic field, $\varphi$ is the magnetic potential, $\alpha$ is the fiber's angle relative to the magnetic field and $H_f$ is the intensity of the magnetic field inside the fiber.}
\label{fig:fiberangle}
\end{figure}

If the fiber is placed in a homogeneous magnetic field with a linear increase in magnetic potential depending on the coordinate, the axial magnetic field intensity within the fiber is determined by the difference in magnetic potentials at its ends (Figure \ref{fig:fiberangle}). If the field direction is parallel to the fiber's axis, the difference in coordinates at the fiber's ends is equal to its length, and the magnetic field intensities within the fiber and the surrounding space are identical. If the field direction is not parallel to the fiber's axis, the difference in coordinates at the fiber's ends is less than its length, depending on the deviation $\alpha$ of its axis from the direction of the field intensity. Using the coordinate system from Figure \ref{fig:fiberangle} we can write

\begin{equation}
    l \cos\alpha = (x_2-x_1)
\end{equation}

The axial intensity of the magnetic field inside the fiber can then be expressed as

\begin{equation}
    H_f l = H_f \frac{x_2 - x_1}{\cos \alpha} = H (x_2 - x_1) \implies H \cos \alpha = H_f
\end{equation}

alternatively as

\begin{equation}
    H_f = \frac{4U}{\pi D^2 \omega \mu N} \cos \alpha
\end{equation}

Power $P_{skin}$ absorbed inside the fiber because of the axial magnetic field can be expressed as follows using the coil's voltage 

\begin{equation}
    P_{skin} = \frac{U_{RMS}^2}{R_d} \cos^2 \alpha
\end{equation}

For the purpose of describing the real dependence of the measuring coil's quality factor on the rotation angle of ferromagnetic fibers, it is necessary to also consider the essentially constant absorbed power, independent of the angle $\alpha$, generated within the fibers in the coil's magnetic field. The total absorbed power can then be expressed as

\begin{equation}
    P_{skin} = \frac{U_{RMS}^2}{R_d} (k_1 \cos^2 \alpha + k_2)
\end{equation}

The damping resistance $R_{ds}$ on the coil's terminals, which is caused by the losses in the ferromagnetic fiber depending on the angle $\alpha$ is then expressed as

\begin{equation}
    R_{ds} = \frac{R_d}{(k_1 \cos^2 \alpha + k_2)}
\end{equation}

And finally, the measured quality factor $Q$ can be written as

\begin{equation}
    Q=\left(\frac{1}{Q_0}+\frac{k_1 \cos^2 \alpha}{Q_M}+\frac{k_2}{Q_M}\right)^{-1}
\label{eq:Q}
\end{equation}

where $Q_0$ is the quality factor without the specimen inside the coil, $Q_M$ is the minimal quality factor caused by the specimen and $k_i$ are approximation constants. In order to approximate $Q$ based on frequency, another equation was chosen with approximation constants to fit the non-linear relationship

\begin{equation}
    Q=\left(\frac{1}{Q_0}+\frac{k_2 + \cos^2 \alpha}{k_3f^3+k_4f^2+k_5f+k_6}\right)^{-1}
\end{equation}

where $f$ is the frequency and the approximation constants are

\begin{equation}
\begin{split}
    k_2 = 0.486638 \\
    k_3 = 0.751142 \\
    k_4 = -11.6495 \\
    k_5 = 49.17971 \\
    k_6 = 46.48844
\end{split}
\end{equation}

\begin{figure}
\centering
\includegraphics[width=0.8\textwidth]{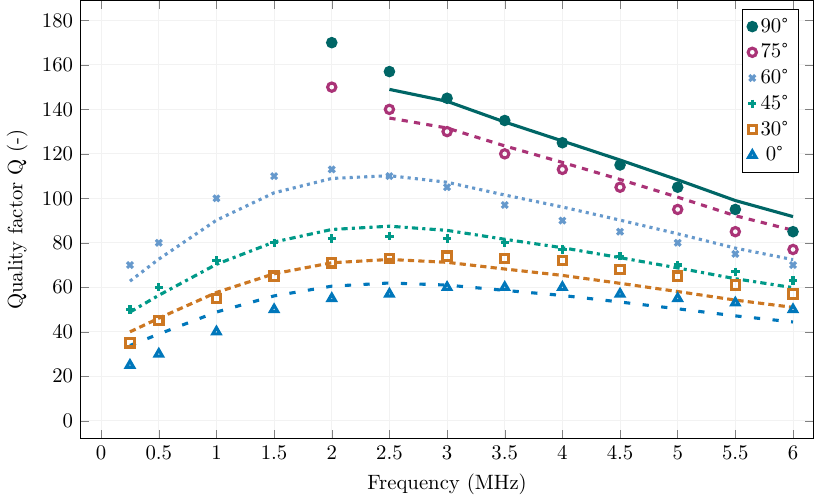}
\caption{Example of influence of frequency and fiber angle on the quality factor. Applicability of the theoretical equation.}
\label{fig:approx}
\end{figure}

This approximation was used to fit a testing measurements of fibers type Dramix 80/30 GCP with volume concentration of 2\%. By choosing the appropriate approximation constants, it is possible to minimize the average error of approximation to around 4\% in the selected frequency range. An example of measured values (markers) and the approximations (lines) are in Figure \ref{fig:approx}.

\subsection{Measuring the quality factor}
\subsubsection{Impedance measurements}

To measure the quality factor, different principles can be used which differ in complexity, accuracy and usability in various environments. The most common method of measuring the quality factor is the impedance approach. The measuring device generates a signal, which it applies to the measured circuit (coil) while measuring vectors of voltage and current. The impedance is then a ratio of the two measured phasors. The quality factor can be determined using the phase deviation $\delta$ of the voltage and current phasors, independently of the measuring frequency as 

\begin{equation}
    Q = \frac{1}{\tan{\delta}}
\end{equation}

This method is usually used by the impedance or LCR meters, which are designed to measure various electrical properties of a circuit in broad frequency range. For specifically measuring the quality factor, the method can be modified by fixing the applied current to a constant value, which makes the impedance directly related to measured voltage phasor. However, this method measures relatively high quality factors of a coil, where the differences caused by ferromagnetic fibers inside the coil are close to negligible. Nevertheless, using a high-quality laboratory impedance meter can still be used in a testing environment to, for example, confirm the functionality of a measuring coil or the whole measuring apparatus and to determine the suitable frequency range for subsequent measurements. This method was used for the first set of laboratory measurements presented in this research (see section \ref{sec:setuplab}).

\begin{figure}
\centering
\includegraphics[width=0.7\textwidth]{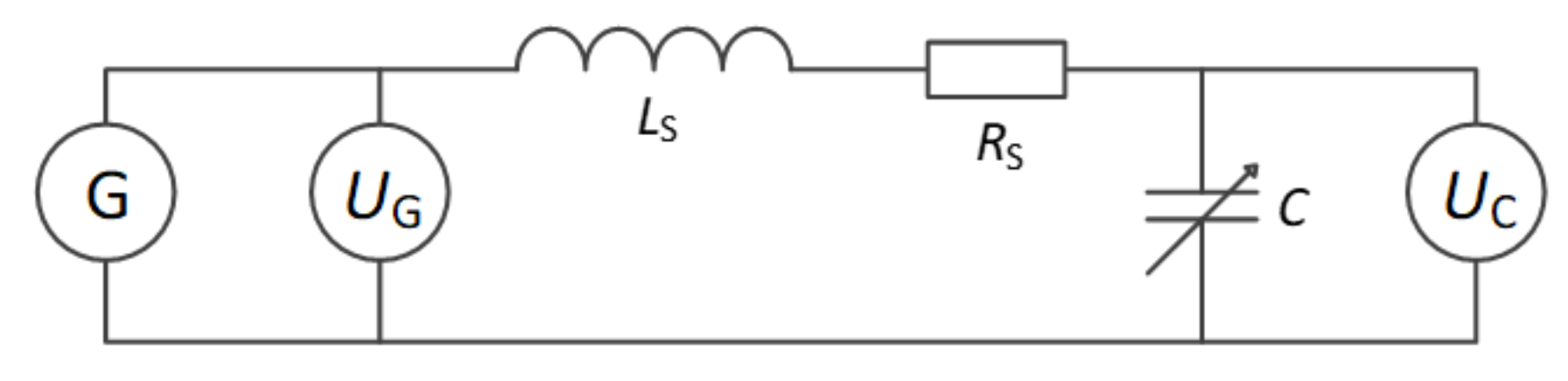}
\caption{Series resonance circuit -- G is the signal generator, $U_G$ is the generator's voltage, $L_S$ is the coil's inductance, $R_S$ is the coil's loss resistance, C is the variable capacitor and $U_C$ is the voltage across the capacitor.}
\label{fig:resonance}
\end{figure}

\subsubsection{Resonance circuit measurements}
\label{sec:resonance}

Another method of measuring a coil's quality factor is based on measuring voltage ratios in a series resonance circuit. This circuit is shown in Figure \ref{fig:resonance}. Capacity of the variable capacitor C is changed during measuring so that together with the coil L there will be resonance at the measuring frequency $f_m$ according to

\begin{equation}
    f_m = \frac{1}{2\pi \sqrt{L_SC}}
\end{equation}

Under this condition, the reactances of the coil L and the capacitor C in a series resonant circuit are compensated in terms of impedance. If a sufficiently high-quality capacitor is used and its losses can be neglected, the circuit's current $I$ is determined exclusively by the output voltage $U_G$ of the generator G and the series damping resistance $R_S$ of the coil L as

\begin{equation}
    I = \frac{U_G}{R_S}
\end{equation}

Voltage $U_C$ across the capacitor C is determined by current $I$ and reactance of capacitor C and is equal to current $I$ and reactance of coil L as

\begin{equation}
    U_C = IX_C = IX_L
\end{equation}

The quality factor $Q$ (see Figure \ref{fig:resonance}) can be expressed as 

\begin{equation}
    Q = \frac{X_{L_S}}{R_S} = \frac{U_C}{U_G}
\end{equation}

The voltage ratio is sometimes described as the magnification factor. 

If the voltage $U_G$ at the output of the measuring signal generator G is kept at a constant level, the quality factor $Q$ is directly proportional to the voltage $U_C$ across the capacitor C. A voltmeter indicating this voltage is often directly calibrated in terms of Q values. In the simplest measuring devices, the constant level of voltage $U_G$ is set manually by the operator based on the voltage indicated by a voltmeter. In high-quality instruments, a constant $U_G$ level is maintained by an automatic regulator. Using a quality factor meter, it is possible to achieve a higher measurement accuracy than in the inductance case, typically around a few percent. This level of measurement accuracy is sufficient to determine changes in the quality factor of the measuring coil due to the ferromagnetic fibers.

However, fibers also cause changes in the coil's inductance. To consistently maintain the resonance condition, this change must be constantly compensated for by adjusting either the resonance capacitance or the measurement frequency. The need for continuous adjustment is the most serious drawback of this measurement procedure, which, without further measures, practically rules out its use for continuous or automated measurement. That is why a specialized measuring circuit was devised as part of a planned industrial meter, which utilities this measuring approach but deals with the aforementioned challenges. The meter is described in section \ref{sec:meter} and was used in the second set of measurements presented in this research. 

\section{Experimental methods and measuring devices}
\subsection{The measuring coils}

\begin{figure}
\centering
\includegraphics[width=0.5\textwidth]{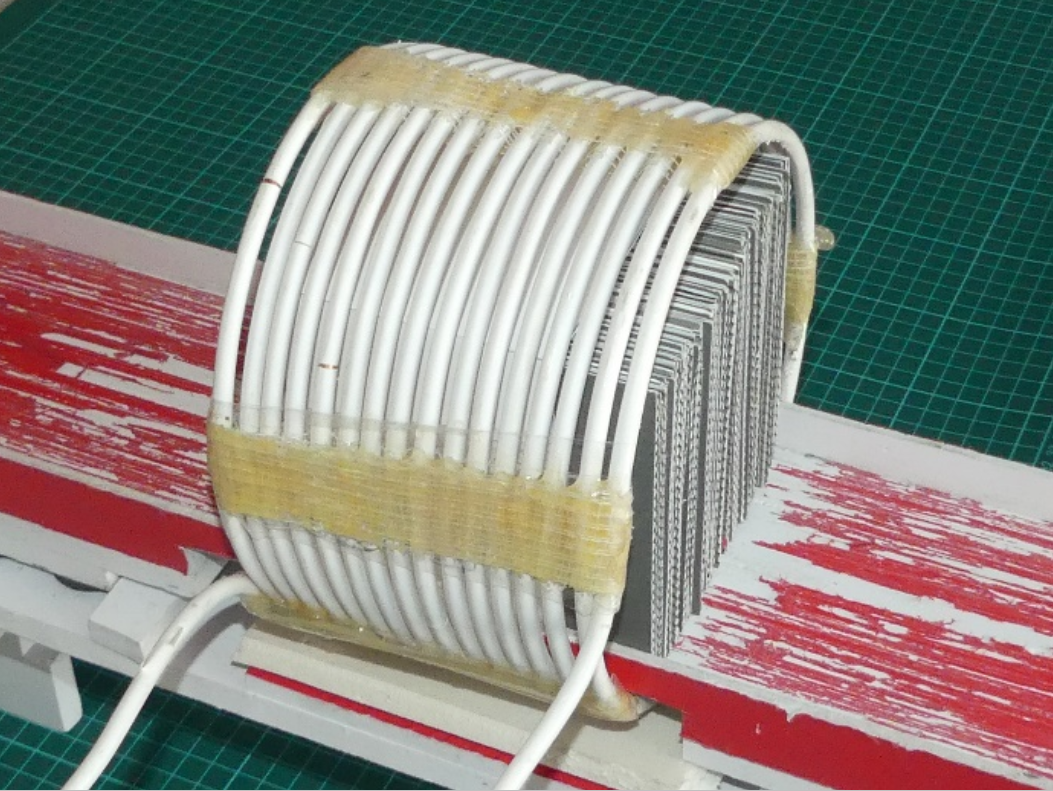}
\caption{Measuring coil for measuring specimens inside the coil. The cardboard specimen is inserted.}
\label{fig:firstcoil}
\end{figure}

Three different coils were used in this study. The first coil (Figure \ref{fig:firstcoil}) was intended to place the entire specimen inside, so it would be possible to rotate it in all axes and asses the orientation influence. This coil, therefore, must have been the air-coil type not suitable for surface measurements, but for the entire volume measurements. The coil was designed for measuring cubic specimens with a side length of 100 mm. It had 16 turns of copper wire with a cross-section of 6 mm$^2$. The mean diameter was 155 mm and length was 110 mm. The coil was fixed using fiberglass fabric saturated with epoxy resin.

\begin{figure}
\centering
\includegraphics[width=0.8\textwidth]{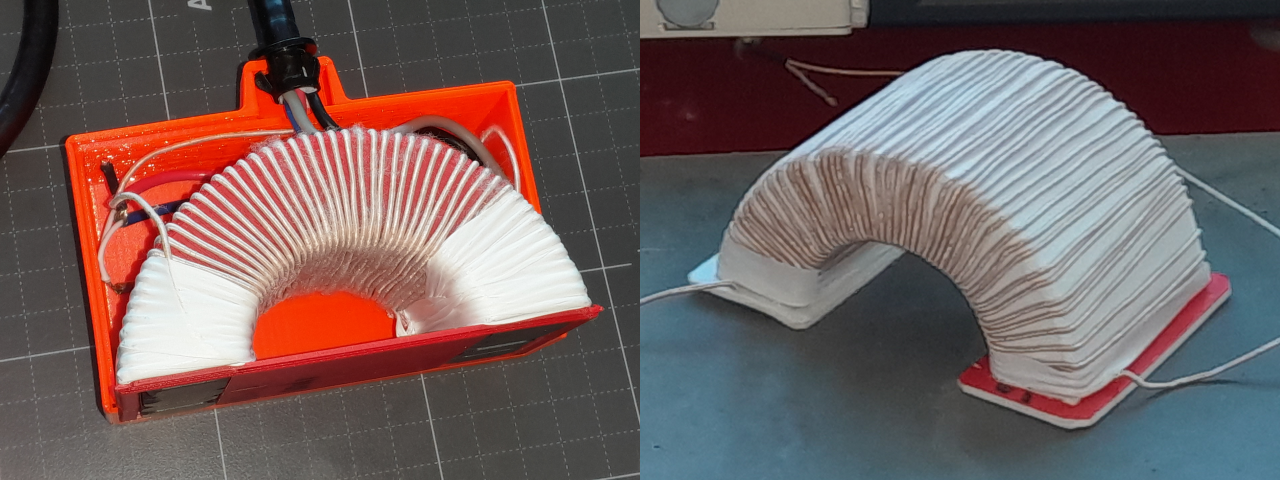}
\caption{The half-toroid measuring coils. Prototype on the left, the improved coil on the right.}
\label{fig:toroid}
\end{figure}

The second and third coils were made for the surface measurements. One was a prototype and the other was an improved version based on the prototype's results. The prototype (Figure \ref{fig:toroid} left) was created using the following procedure: The core was made by cutting a toroidal core with a rectangular cross-section from K10 material in half using a water jet. It had an outer diameter of 87 mm, an inner diameter of 54.3 mm, and a thickness of 13.5 mm. A frame made of foamed PVC was built around it, and 50 turns of high-frequency stranded wire were wound. This wire consisted of 120 conductors, each 0.1 mm in diameter, braided with silk insulation. During construction, it was essential to minimize the parasitic parameters of the measuring coil. This ensures that its influence on the measurement results would be less significant compared to the influence of the fibers being tested. 

\begin{figure}
\centering
\includegraphics[width=0.8\textwidth]{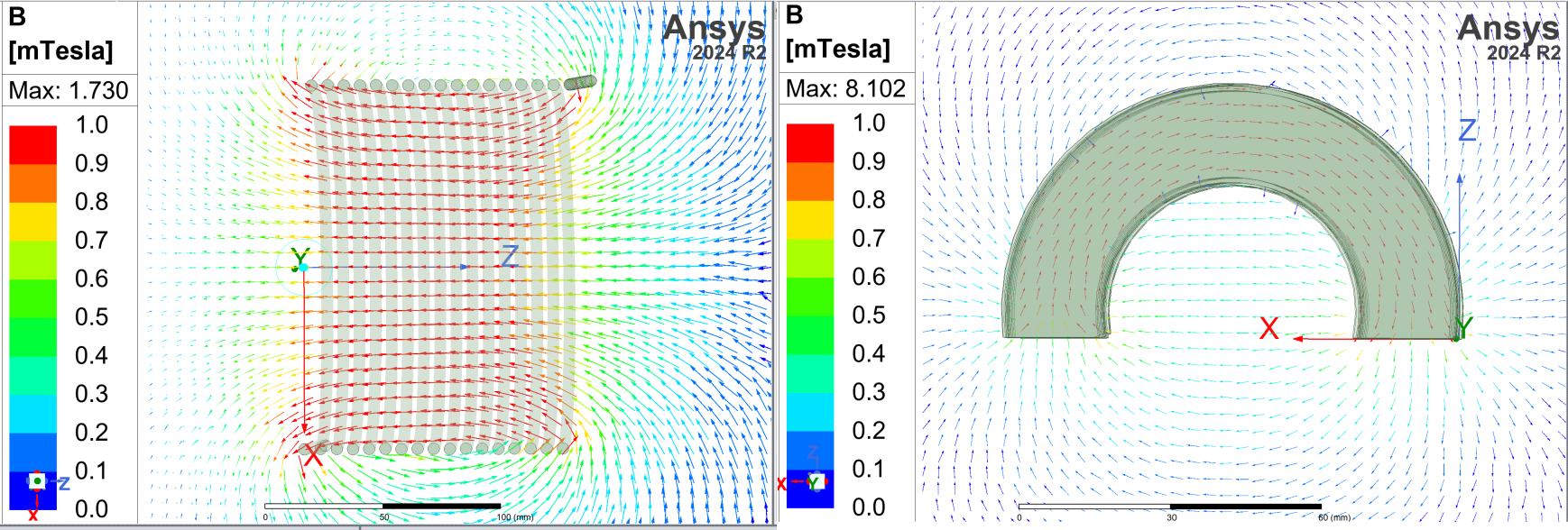}
\caption{Differences in magnetic fields' shapes between the hollow circular (left) and toroidal core (right).}
\label{fig:sim}
\end{figure}

The final surface-measuring coil (Figure \ref{fig:toroid} right) was made larger, to enhance the measured differences of quality factors with different fiber conditions. The new core was made using four half-toroid cores and the same wiring as the prototype. Figure \ref{fig:sim} shows the difference in magnetic field created by the air coil and a half-toroid coil. It was important to chose large enough toroid core to have sufficient distance between its free ends, to maximize the measured volume, but also to minimize the transition area between the end and the specimen. The shapes of the magnetic fields' for measuring and orienting the fibers during manufacturing are practically identical, with the measuring field's intensity being several orders of magnitude lower, but measuring is conducted at much higher frequency. 

It must be stressed, that each coil was specifically constructed for the quality factor measuring principle. This means, that the coil's quality factor was as high as possible, to maximize their sensitivity to a ferromagnetic material placed inside them or in their vicinity. It should also be noted that the quality factor values alone are meaningless, as they are always tied to the specific measuring coil and measuring frequency. The measured values must always be compared for different amounts, or orientations, of fibers.

\subsection{Design of the surface measuring device}
\label{sec:meter}

A quality factor meter operating on the resonant principle (see Section \ref{sec:resonance}) was designed, developed, and tested for the purpose of nondestructive diagnostics. The measuring frequency was chosen as the compensation parameter because it can be controlled electronically with relative ease. This isn't the case with resonant capacitance.

\begin{figure}
\centering
\includegraphics[width=1\textwidth]{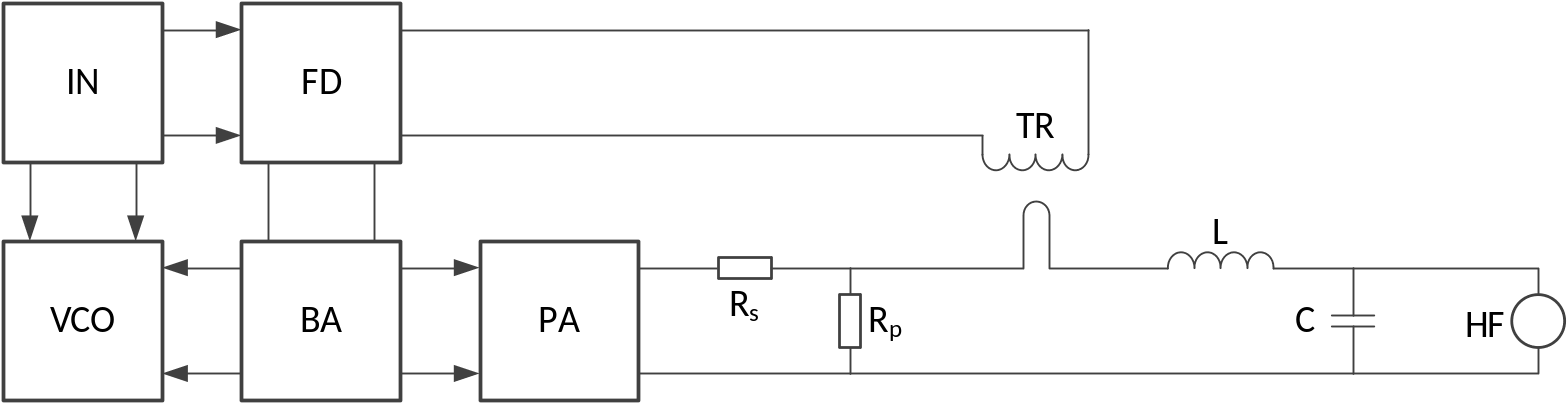}
\caption{Circuit of the quality factor meter.}
\label{fig:circuit}
\end{figure}

Figure \ref{fig:circuit} shows the quality factor meter's circuit. The measuring coil L is connected by one terminal to the resonant capacitor C, to which a high-frequency voltmeter HF is connected in parallel to indicate the oscillating voltage. To achieve minimal damping of the resonant circuit by the indicating voltmeter, the voltmeter is designed as a special peak detector with a very high input impedance. This impedance exceeds the reactance of the resonant capacitor C by approximately two orders of magnitude in the range of measuring frequencies.

The variable frequency measuring signal is generated by a voltage-controlled oscillator VCO, whose oscillation frequency is controlled by the phase-locked loop, a driver BA that minimizes the influence of connected circuits, and a power amplifier PA. The PA is designed as a push-pull switch, which generates a symmetrical rectangular voltage waveform at its output. The use of a pulsed excitation signal is possible due to the high selectivity of the measuring circuit. Since the series resonant circuit in this arrangement functions as a highly selective band-pass filter, its response to a pulsed excitation signal is determined practically only by the first harmonic component of its spectrum, which corresponds to its resonant frequency. At the same time, the switch exhibits minimal internal resistance, ensuring that the level of its output voltage is minimally dependent on the load of the measuring circuit.

The mutual influence of the PA and the measuring circuit is further reduced by a resistive divider, R$_S$ and R$_P$, which sets the voltage supplied to the measuring coil to approximately 0.1 V with an internal impedance of the measuring signal source of less than 0.2 $\Omega$. The feedback loop, consisting of the VCO, BA, PA, LC resonant circuit, current transformer TR, phase detector FD, and integrator IN, forms the phase-locked loop control system. This system compares the phase of the LC resonant circuit's excitation voltage with the phase of the current flowing through it and adjusts the frequency of the VCO so that the instantaneous phase deviation is minimal. This automatically ensures that the resonant condition is continuously maintained in the LC circuit during measurement. The quality factor of the coil L then corresponds to the voltage measured by HF.

\begin{figure}
\centering
\includegraphics[width=0.6\textwidth]{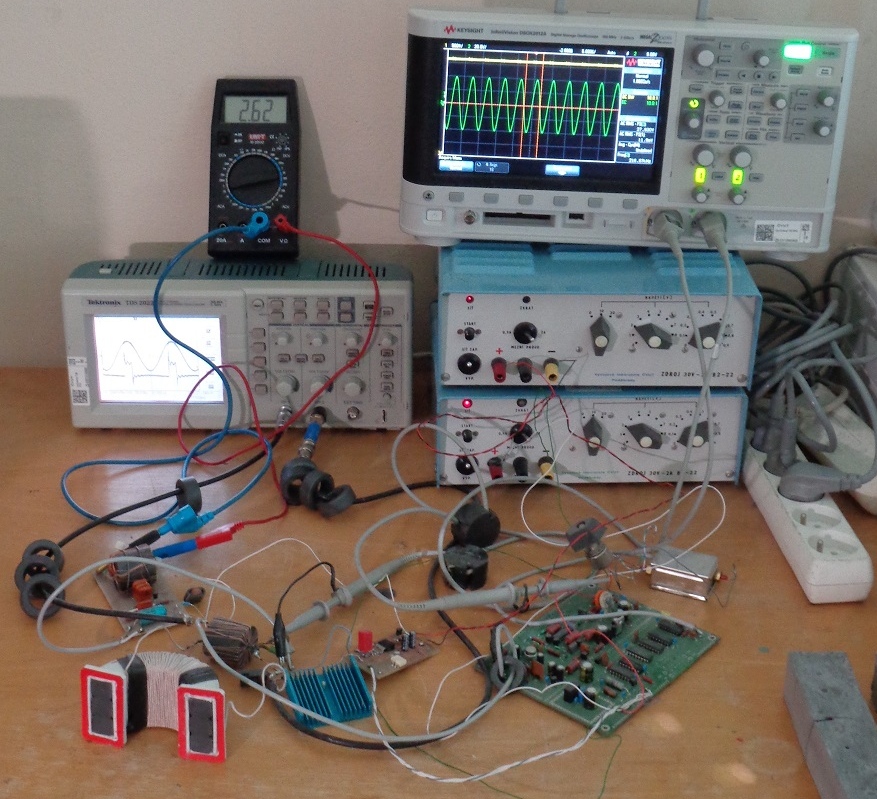}
\caption{Prototype of the quality meter. Note the attached half-toroid core.}
\label{fig:circuit2}
\end{figure}

The feedback branch of the phase-locked loop consists of the measuring current transformer TR, the phase detector FD, and the integrator IN. The output signal from the secondary winding of the measuring current transformer TR, corresponding to the current sample in the measuring circuit, is fed to the first input of the phase detector FD. A comparison signal, representing the voltage on the measuring circuit, is fed to its second input from the driver BA. The signal from the output of the phase detector FD is sent to the integrator IN, which acts as a low-pass filter for the phase-locked loop. The signal from the integrator's output is the control signal that is then fed to the control input of the VCO. The device is protected by Czech and European patents \cite{CZ309919B6}, \cite{EP4386367B1} and a prototype laboratory assembly is in Figure \ref{fig:circuit2}.

\subsection{Laboratory specimens}

\begin{figure}
\centering
\includegraphics[width=1\textwidth]{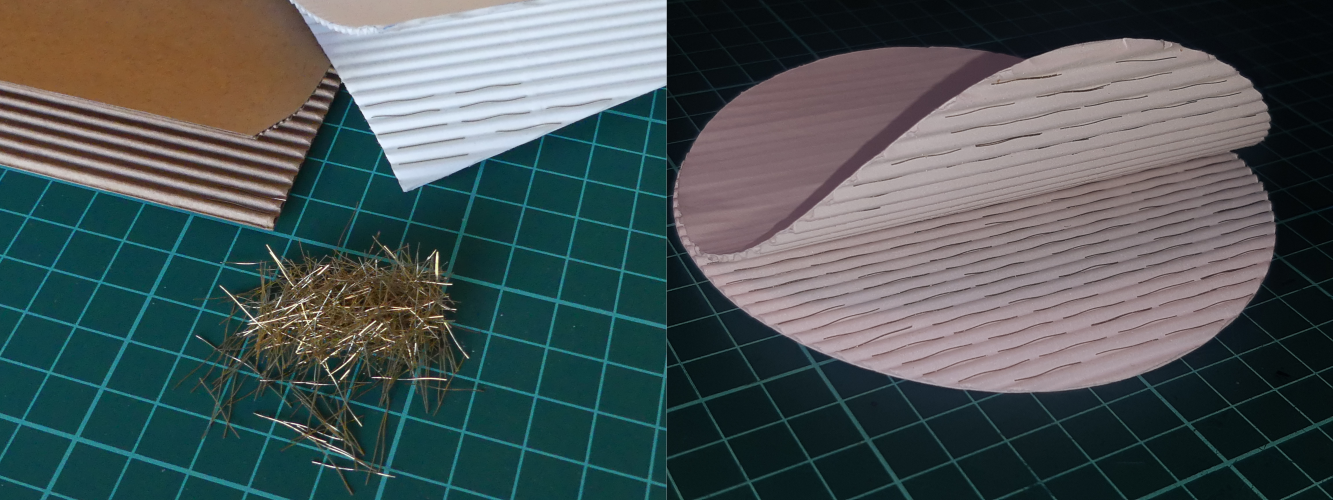}
\caption{fiber specimens in cardboard carrier. Rectangular (left) and circular (right).}
\label{fig:cardboard}
\end{figure}

For the experimental campaign, specimens needed to be prepared with exactly the specified amount and positions of steel fibers. Fibers chosen were cylindrical, straight, steel Weidacon FM, 13 mm long and 0.19 mm in diameter, as these and similar fibers are commonly used in high-performance fiber-reinforced concrete, which is the main focus of the authors' research. Corrugated cardboard was used as a carrier for fibers. For the circular coil a cube specimen was needed. This was done using 50 sheets of cardboard 100 mm $\times$ 100 mm $\times$ 2 mm. The sheet was split to insert the fibers and then closed with adhesive. fibers took 0.35 \% of the sheet's volume. Different sheets were prepared where the corrugations were aligned at 0\textdegree, 15\textdegree, 30\textdegree and 45\textdegree. To measure the influence of concentration, sheets without fibers were also prepared. For the toroidal surface measurements, circular sheets were prepared with 124 mm in diameter with 2\% of volume of fibers. The specimens are in Figure \ref{fig:cardboard}, showing how the cardboard keeps the manually-placed fibers constantly oriented with the corrugation.  

\newpage

\subsection{The experimental setup}
\subsubsection{Air coil with cube specimen - induction measurements}
\label{sec:setuplab}

For the air coil, 50 cardboard sheets were wrapped in paper to form a cubic specimen. The number of sheets with fibers was varied in steps of 5 from 15 to 50. The rest of the specimen was filled with sheets without fibers. Fibers in the specimen always shared the same orientation. The orientations used were 0\textdegree, 15\textdegree, 30\textdegree and 45\textdegree. The cube specimen fit entirely into the measuring coil, therefore, it could be inserted in three different orientations, i.e. the specimen can be measured in three axes. In the evaluation, the aforementioned theoretical relationships were utilized, where it was assumed that the quality factors in individual axes $Q_X$, $Q_Y$ and $Q_Z$ are summed as their reciprocal values. The no-load quality factor is subtracted from the partial measurements, so neglecting the cosine dependence on fiber rotation, we can determine the overall quality factor $Q$ as

\begin{equation}
\label{eq:sumq}
    \frac{1}{Q} = \frac{1}{Q_x} + \frac{1}{Q_y} + \frac{1}{Q_z} - \frac{3}{Q_0}
\end{equation}

Another experiment was aimed at confirming the concentration assumptions. All sheets contained fibers this time, but sheets in steps of 5 from 15 to 50 contained fibers oriented perpendicularly to the rest of the sheets. All experiments with the aircoil were done using the impedance principle, by connecting the measuring coil to a handheld impedance meter GW Instek LCR-1100, which uses a fixed measuring frequency of 100 kHz. 

\subsubsection{Half-toroid coil with circular specimens - induction and the specialized meter measurements}

Both of the introduced half-toroid coils were used in the experiments, as the improved version was created as a result of pilot tests with the prototype. A stack of the circular cardboard sheets was used as the measured specimen. Up to 11 sheets with fibers were used. Two sets of measurements were conducted -- focused on fiber content and fiber orientation. For fiber content, the number of uniformly oriented sheets was changed between 0; 1; 2; 4; 8 and 11. Similarly for orientation, a fixed number of 10 sheets were turned against the measuring coil at 0\textdegree, 30\textdegree, 60\textdegree and 90\textdegree. This measurement needed to also study the effect of frequency. The frequency range was chosen from 100 Hz to 1 MHz, therefore the desktop LCR meter HIOKI IM3536 was used. 

Using the improved half-toroid coil, experiments were done to verify the design of the quality factor meter intended for industrial environment. New set of measurements were done using 0 to 10 circular sheets of fibers, oriented at 0\textdegree and 90\textdegree relative to the coil. This set was measured using all three available devices -- the novel prototype quality factor meter set to 217 kHz, the HIOKI desktop LCR meter set to 215 kHz (closest to the prototype meter) and the GW Instek handheld LCR meter able to measure at its fixed frequency of 100 kHz.

\section{Results and discussion}

\subsection{Air coil measurements}

\begin{figure}[h!]
\centering
\begin{subfigure}[t]{0.48\textwidth}
\centering
\includegraphics[width=\textwidth]{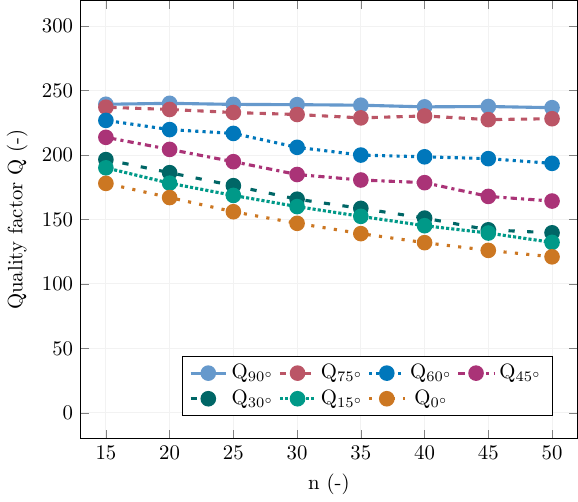}
\caption{}
\end{subfigure}\hfill
\begin{subfigure}[t]{0.48\textwidth}
\centering
\includegraphics[width=\textwidth]{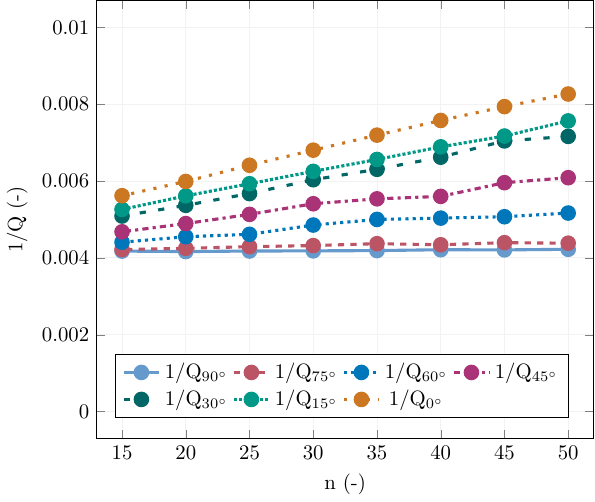}
\caption{}
\end{subfigure}
\caption{Example of air coil one-axis measurements only, (a) quality factor, (b) reciprocal quality factor.}
\label{fig:air1}
\end{figure}

Figure \ref{fig:air1} shows an example of results $Q_X$ where the cardboard cube was inserted in x axis only with varying numbers of sheets filled with fibers. If all of the fibers were oriented perpendicular to the air coil, the amount of fibers exhibited negligible change on the coil's performance. On the other hand, fibers oriented parallel to the coil's axis changed its quality factor the most, as expected. Figure \ref{fig:air1}b shows the reciprocal values. 

\begin{figure}[h!]
\centering
\begin{subfigure}[t]{0.48\textwidth}
\centering
\includegraphics[width=\textwidth]{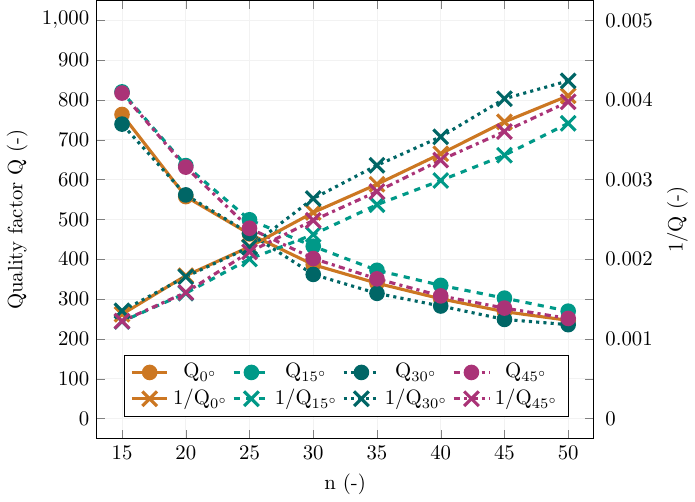}
\caption{}
\end{subfigure}\hfill
\begin{subfigure}[t]{0.48\textwidth}
\centering
\includegraphics[width=\textwidth]{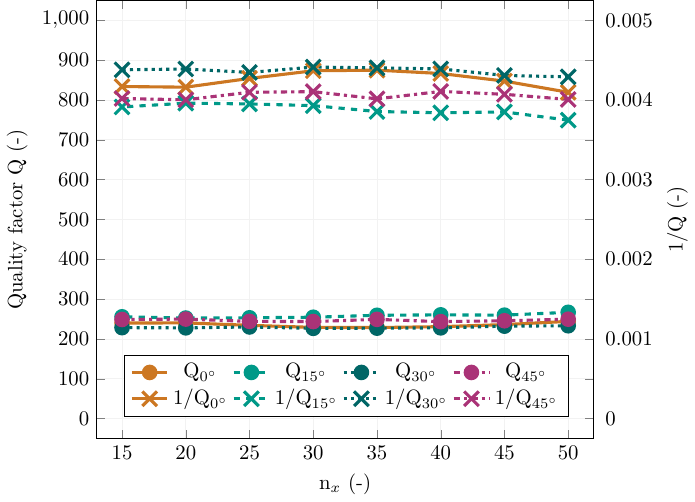}
\caption{}
\end{subfigure}
\caption{Summed quality factor with the air coil, (a) effect of content and orientation, (b) effect of orientation only.}
\label{fig:air2}
\end{figure}

Figure \ref{fig:air2}a shows the resulting quality factor sum as outlined by Equation \ref{eq:sumq} (both $Q$ and $1/Q$) when the amount of fibers gradually increased. Only values from 0\textdegree to 45\textdegree are shown, as those are the angles towards the x axis, meaning that, for example 60\textdegree results would be identical to 30\textdegree, as it would only switch orientations between x and y axes. It can be seen, that the amount of fibers clearly influences the resulting quality factor. The relative similarity between the orientations is acceptable and confirms the theoretical assumptions indicated by Equation \ref{eq:sumq}. Figure \ref{fig:air2} shows the results for the situation when the amount if fibers was kept constant, but orientations were changed, as described in Section \ref{sec:setuplab}. Again, the theoretical assumptions were confirmed, as the sum of partial reciprocal quality factors measured in 3 axes negate the influence of orientation and allow the study of fiber content only.

\subsection{Half-toroid coil measurements for surface fiber content and orientation}

\begin{figure}[h!]
\centering
\begin{subfigure}[t]{0.48\textwidth}
\centering
\includegraphics[width=\textwidth]{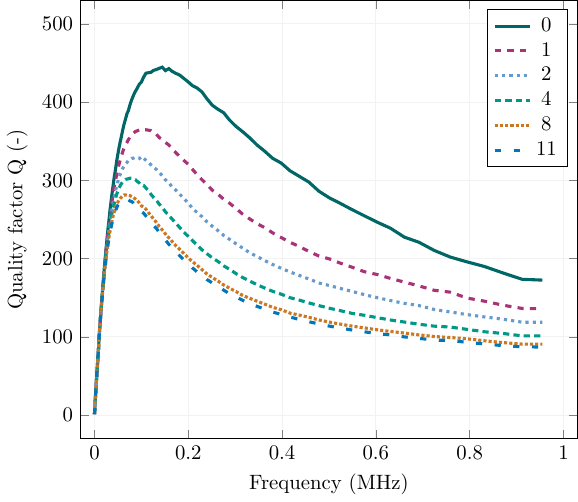}
\caption{}
\end{subfigure}\hfill
\begin{subfigure}[t]{0.48\textwidth}
\centering
\includegraphics[width=\textwidth]{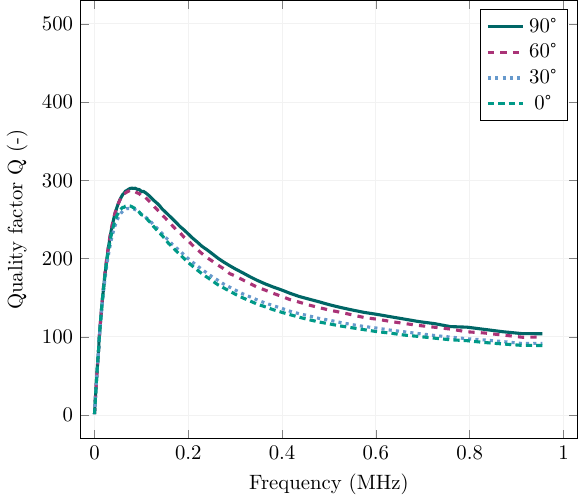}
\caption{}
\end{subfigure}
\caption{Results for the prototype half-toroid coil, (a) fiber content study, (b) fiber orientation study.}
\label{fig:surface1}
\end{figure}

Figure \ref{fig:surface1} shows the results using the first prototype of the half-toroid measuring coil. Figure \ref{fig:surface1}a shows the influence of adding uniformly oriented fiber sheets. It can be seen, that there was a significant decrease in quality factor between 0 and 1 sheet. Each additional sheet created lower difference when finally 8 and 11 sheets were practically indistinguishable. This shows the penetration ability of the measuring principle. Figure \ref{fig:surface1}b shows the influence of differently oriented 10 sheets relative to the measuring coil. In this particular case, it is clear that the method exhibited poor sensitivity.  

\begin{figure}[h!]
\centering
\begin{subfigure}[t]{0.48\textwidth}
\centering
\includegraphics[width=\textwidth]{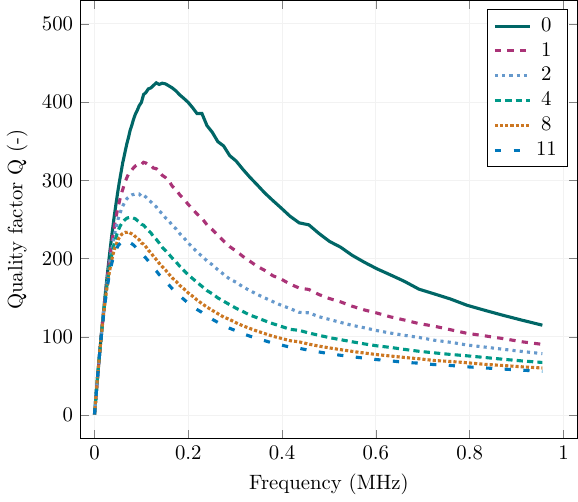}
\caption{}
\end{subfigure}\hfill
\begin{subfigure}[t]{0.48\textwidth}
\centering
\includegraphics[width=\textwidth]{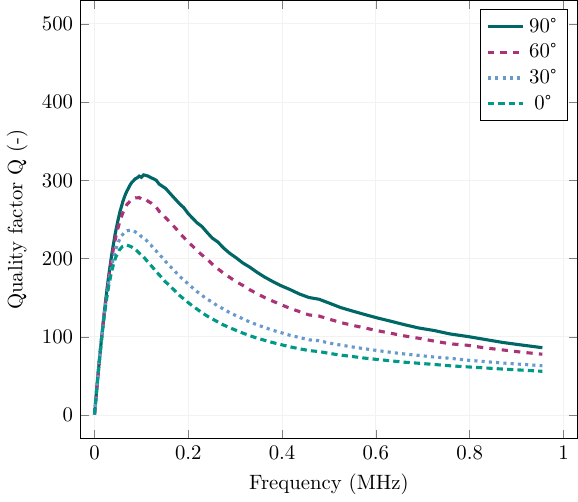}
\caption{}
\end{subfigure}
\caption{Results for the improved half-toroid coil, (a) fiber content study, (b) fiber orientation study.}
\label{fig:surface2}
\end{figure}

These results prompted the creation of the second half-toroid coil with increased measuring area and therefore, sensitivity. All of the results from now on were obtained using this improved half-toroid core. Figure \ref{fig:surface2}a once again shows the influence of adding uniformly oriented fiber sheets. The differences are similar as before, with slightly better resolution for higher number of sheets. Figure \ref{fig:surface2}b shows the orientation data, where the differently oriented sheets now exhibit clearly different quality factor values. Overall it can also be seen that the highest sensitivities for fiber concentrations and orientations are achieved in a frequency range from approximately 80 kHz to 300 kHz.

\begin{figure}[h!]
\centering
\begin{subfigure}[t]{0.48\textwidth}
\centering
\includegraphics[width=\textwidth]{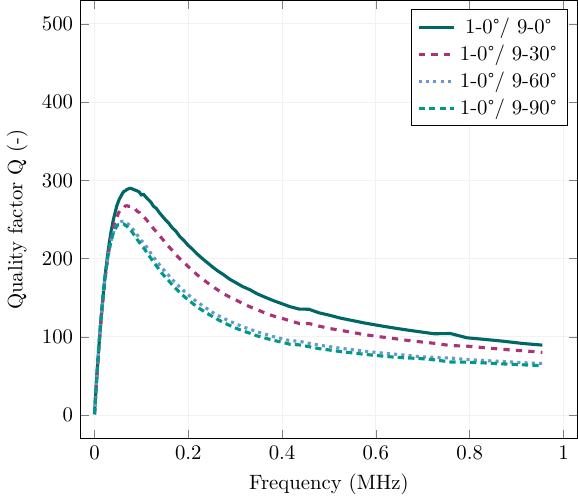}
\caption{}
\end{subfigure}\hfill
\begin{subfigure}[t]{0.48\textwidth}
\centering
\includegraphics[width=\textwidth]{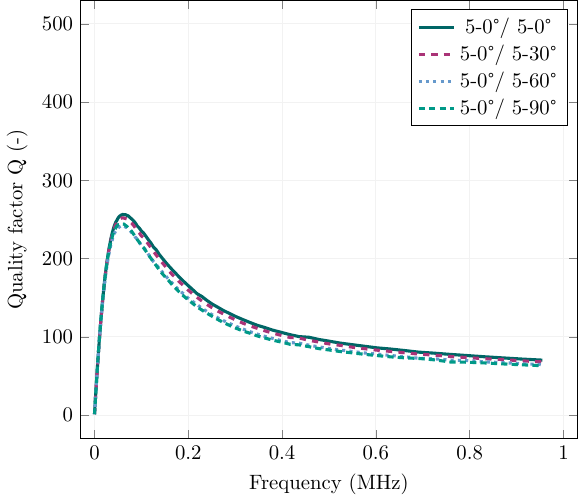}
\caption{}
\end{subfigure}
\caption{Fiber shielding effect with the improved half-toroid coil, (a) first layer constant, (b) first 5 layers constant.}
\label{fig:surface3}
\end{figure}

Figures \ref{fig:surface3} shows results of experiments, where the shielding effect of the surface layers affects measuring the deeper layers. Figure \ref{fig:surface3}a was a situation, where the first layer (closest to the coil) was oriented at 0\textdegree, while 9 other layers below it were oriented according to the chart's legend. It can be seen, that the measurements are significantly less sensitive, compared to the uniform orientations before. Figure \ref{fig:surface3}b is then a situation when 5 layers closest to the surface have fixed 0\textdegree orientation, while the other 5 layers change orientation. In this case, the orientations are practically indistinguishable from each other. This confirms, that the shielding effect is quite strong and must be taken into consideration when measuring relatively thick specimens. In practical applications, it would be beneficial if a concrete volume was measured from both sides. Nevertheless, if measuring an element that would undergo tension in the measured area, the surface layers are of great importance, therefore the measuring approach would be able to predict the mechanical behavior despite the shielding effect. 

\subsection{Testing the prototype quality factor industrial meter}

\begin{figure}[h!]
\centering
\begin{subfigure}[t]{0.48\textwidth}
\centering
\includegraphics[width=\textwidth]{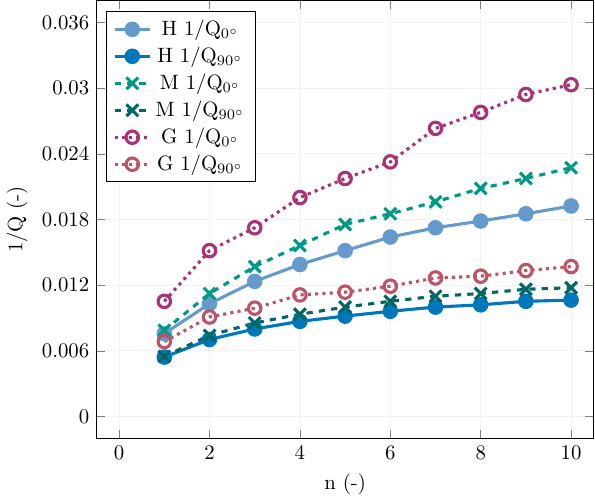}
\caption{}
\end{subfigure}\hfill
\begin{subfigure}[t]{0.48\textwidth}
\centering
\includegraphics[width=\textwidth]{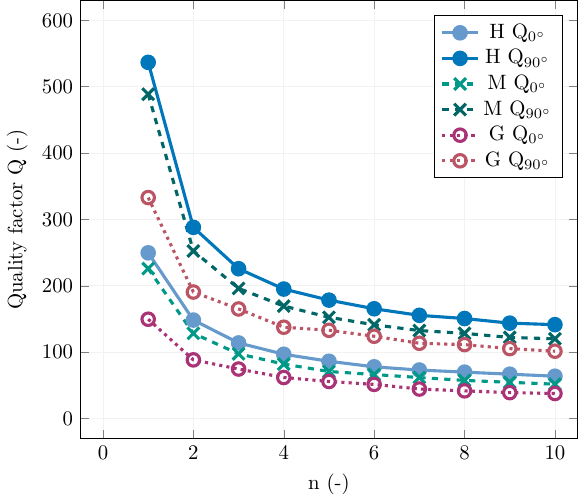}
\caption{}
\end{subfigure}
\caption{Comparison of the 3 measuring devices, (a) reciprocal quality factor, (b) quality factor.}
\label{fig:meter}
\end{figure}

The final results were used to analyze the suitability of the newly developed quality factor meter. Figure \ref{fig:meter} shows the comparison of the new meter (M) with the commercial meters HIOKI (H) and GW Instek (I) for different amount of fiber cardboard discs oriented at 0\textdegree and 90\textdegree. Figure \ref{fig:meter}a shows the reciprocal values with the no-fiber reciprocal measurement subtracted. Figure \ref{fig:meter}b then shows the final quality factor values. As the quality factor measurements are comparative in principle, the individual values are meaningless for these purposes. However, it can be seen, that the trends of all lines are practically identical, which confirms the correct measurements of the newly developed meter. The sensitivity to the fiber orientation is sufficient, meaning that the oriented states can be clearly distinguished. 

\section{Conclusion}

This research presented the theoretical background and experimental confirmation of using the coil's quality factor for determining the amount and orientation of steel fibers in fiber-reinforced concrete. It was confirmed the the theoretical assumptions were correct. The air-coil approach was upgraded using a half-toroid core to be able to measure the concrete specimen from one side only, which would allow the analysis of practically infinitely long and wide specimens. Moreover, a novel quality-factor meter was developed, presented and tested, which removes the necessity of using a costly and sensitive laboratory LCR meter. This new meter is suitable for applications in the prefabrication industry, for final quality control of the manufactured products. However, the non-destructive approach does have its limits. The surface measurements are mostly influenced by the fiber layers closer to the surface. Their parameters shield the layers further inside, which could limit the usability on thicker specimens. The quality factor values themselves are also practically meaningless, as the measurement principle is comparative and must be conducted on relatively high number of points on the specimen. That is likely not a problem in an automated production line. Overall the quality factor analysis principle is suitable to determine the fiber parameters with sufficient accuracy. 

\bibliographystyle{bibstyle.bst}
\bibliography{biblio}

\begin{thebibliography}{10}
\providecommand{\url}[1]{\texttt{#1}}
\providecommand{\urlprefix}{URL }
\providecommand{\eprint}[2][]{\url{#2}}
\makeatletter
\def\bibdoi{\begingroup\def\do##1{\catcode `##112\relax}\do$\do\\\do\_\do\%\do\^\expandafter\endgroup\@bibdoi}
\def\@bibdoi#1{\eprint{https://doi.org/#1}}
\makeatother

\bibitem{Miller2020}
S.~A. Miller, F.~C. Moore.
\newblock Climate and health damages from global concrete production.
\newblock \textit{Nature Geoscience} 2020.
\bibdoi{10.1038/s41558-020-0733-0}
\bibitem{Mosley2007}
W.~H. Mosley, J.~H. Bungey, R.~Hulse.
\newblock \textit{Reinforced Concrete Design}.
\newblock Palgrave Macmillan, New York, NY, 6th edn., 2007.

\bibitem{Bentur2007}
A.~Bentur, S.~Mindess.
\newblock \textit{Fibre reinforced cementitious composites}.
\newblock Taylor \& Francis, London, UK, 2nd edn., 2006.

\bibitem{DiPrisco2009}
M.~Di~Prisco, G.~Plizzari, L.~Vandewalle.
\newblock Fibre reinforced concrete: new design perspectives.
\newblock \textit{Materials and Structures} \textbf{42}(9):1261--1281, 2009.
\bibdoi{10.1617/s11527-009-9529-4}
\bibitem{BOULEKBACHE20101664}
B.~Boulekbache, M.~Hamrat, M.~Chemrouk, S.~Amziane.
\newblock Flowability of fibre-reinforced concrete and its effect on the mechanical properties of the material.
\newblock \textit{Construction and Building Materials} \textbf{24}(9):1664--1671, 2010.
\bibdoi{https://doi.org/10.1016/j.conbuildmat.2010.02.025}
\bibitem{WIJFFELS2017342}
M.~Wijffels, R.~Wolfs, A.~Suiker, T.~Salet.
\newblock Magnetic orientation of steel fibres in self-compacting concrete beams: Effect on failure behaviour.
\newblock \textit{Cement and Concrete Composites} \textbf{80}:342--355, 2017.
\bibdoi{https://doi.org/10.1016/j.cemconcomp.2017.04.005}
\bibitem{Ghailan02012020}
D.~B. Ghailan, A.~A. Al-Ghalib.
\newblock Magnetic alignment of steel fibres in self-compacting concrete.
\newblock \textit{Australian Journal of Structural Engineering} \textbf{21}(1):333--341, 2020.
\bibdoi{10.1080/13287982.2019.1643642}
\bibitem{ALRIFAI2024134796}
M.~M. {Al Rifai}, K.~S. Sikora, M.~N. Hadi.
\newblock Magnetic alignment of micro steel fibers embedded in self-compacting concrete.
\newblock \textit{Construction and Building Materials} \textbf{412}:134796, 2024.
\bibdoi{https://doi.org/10.1016/j.conbuildmat.2023.134796}
\bibitem{cryst11070837}
R.~Mu, R.~Dong, H.~Liu, et~al.
\newblock Preparation of aligned steel-fiber-reinforced concrete using a magnetic field created by the assembly of magnetic pieces.
\newblock \textit{Crystals} \textbf{11}(7), 2021.
\bibdoi{10.3390/cryst11070837}
\bibitem{CARRERA2023104534}
K.~Carrera, K.~Künzel, P.~Konrád, et~al.
\newblock The effect of magnetic field parameters on fibre orientation in high-performance fibre-reinforced concrete.
\newblock \textit{Mechanics of Materials} \textbf{177}:104534, 2023.
\bibdoi{https://doi.org/10.1016/j.mechmat.2022.104534}
\bibitem{acta1}
K.~Carrera, K.~Künzel, P.~Konrád, et~al.
\newblock Concrete lintels reinforced with steel fibres oriented by a magnetic field.
\newblock \textit{Acta Polytechnica} \textbf{62}(5):531–537, 2022.
\bibdoi{10.14311/AP.2022.62.0531}
\bibitem{doi:10.1061-ASCE-MT.1943-5533.0004342}
K.~Carrera, K.~Künzel, P.~Konrád, et~al.
\newblock Mechanical properties of high-performance concrete with steel fibers oriented by an electromagnetic field.
\newblock \textit{Journal of Materials in Civil Engineering} \textbf{34}(9):04022199, 2022.
\bibdoi{10.1061/(ASCE)MT.1943-5533.0004342}
\bibitem{Ferrara2012}
L.~Ferrara, M.~Faifer, S.~Toscani.
\newblock A magnetic method for non destructive monitoring of fiber dispersion and orientation in steel fiber reinforced cementitious composites--part 1: Method calibration.
\newblock \textit{Materials and Structures} \textbf{45}(4):575--589, 2012.
\bibdoi{10.1617/s11527-011-9793-y}
\bibitem{CZ309919B6}
V.~Papež, K.~Künzel.
\newblock Zařízení pro diagnostiku rozptýlené výztuže v cementovém kompozitu, 2023.

\bibitem{EP4386367B1}
V.~Papež, K.~Künzel.
\newblock Apparatus for diagnostics of dispersed reinforcement in cementitious composite, 2024.

\end{thebibliography}

\end{document}